\documentclass[aps,a4paper,
twocolumn,
rpsfig]{revtex4}

\usepackage{bm}
\usepackage{mathrsfs}
\usepackage{xcolor,color,graphicx,graphics}
\usepackage[all]{xy}
\usepackage{epsfig,subfigure}
\usepackage{latexsym,amssymb,amsmath,amsfonts} 
\usepackage[english]{babel} 
\usepackage[OT1]{fontenc}
\usepackage[latin1]{inputenc}
\usepackage{makeidx}
\usepackage{hyperref}
\usepackage{color,graphicx,graphics,wrapfig,epsf}

\begin{document}

\title{Possible formation of ring galaxies  by  torus - shaped magnetic wormholes }
\author{A.A. Kirillov}
\email{ka98@mail.ru}
\affiliation{Bauman Moscow State Technical University, Moscow, 105005, Russian Federation}
\author{E.P. Savelova}
\email{savelova@bmstu.ru}
\affiliation{Bauman Moscow State Technical University, Moscow, 105005, Russian Federation}


\begin{abstract}

We present the hypothesis that 
some of ring galaxies 
were formed by relic magnetic torus - shaped wormholes.  In the primordial plasma before the
recombination magnetic fields of wormholes trap baryons whose energy is smaller than a threshold energy. They work as the Maxwell's demons 
collecting baryons from the nearest (horizon size) region and thus forming clumps of baryonic matter which have the same torus-like shapes as wormhole throats.
Such clumps may serve as seeds for the  formation of  ring galaxies and smaller objects having the ring form. 
Upon the recombination torus-like clumps may decay and merge. Unlike galaxies, such objects may contain less or even no dark matter in halos. However the most stringent feature of such objects is the presence of a large - scale toroidal magnetic field.
We show that there are threshold values of magnetic fields  which give
 the upper and lower boundary values for the  baryon clumps in such  protogalaxies. 
 

\end{abstract}
\maketitle

\section{Introduction}
The widely accepted theory of galaxy formation is based on Lambda-CDM model. The primordial inhomogeneities in dark matter develop and first form relatively small clumps of matter which eventually merge and form galaxies and groups. This picture represents the hierarchical process  \cite{Hierarchy1,Hierarchy2} and the theory is heavily based on the presence of sufficiently strong inhomogeneities in cold dark matter ($\delta \rho _{DM} / \rho _{DM}\sim 10^{-3}$ at the moment of the recombination). It is rather successful and only slightly underestimates the number of thin disk galaxies in the universe. It is recognized that upon some refining it allows to reproduce the population of galaxies in our Universe.

Some part of galaxies are shaped like a doughnut, the so-called ring galaxies \cite{R1,R2}, and may contain a hole in the middle, e.g., see the recently found giant ring galaxy (ID 5519) \cite{R3}.  In fact, more than a half of disk galaxies have ring structures of different scales \cite{R2} which may also have an irregular form \cite{R4}.
A large amount of the ring galaxies are accompanied by galaxies which are situated near the center of the ring. There are a lot
of works which describe the formation of a ring after the inner galaxy. Firstly, it is necessary to mention the model of formation
which is connected with the bar in the galaxy and describes the non-axisymmetric perturbations of the gravitational potential. The 
bar rotation is connected with Lindblad resonance which causes the radial perturbations of density. This approach was described by
Buta and Combes \cite{R1} and further development can be found in the work of Romero-Gomez
et al. \cite{R5}. Another mechanism which is quite popular in astronomical papers is connected with the fall of the satellite object which
causes the formation of the ring. Such approach is developed, for example, by Wu and Jiang \cite {R6}. Also there are another models which may explain 
the origin of outer stellar rings by the accretion of a cold gas from outside of the galaxy, e.g., see discussions in \cite{R7,R8} and references therein.

However, some of such galaxies have neither  an accompanying galaxy to be produced by the impact process, nor a bar to have the resonance nature. 
The mechanism based on the accretion of a gas  works in this case.
Nevertheless, the ideal form of the ring in some galaxies, e.g., the Hoag's object \cite{Hoag}, or of the above mentioned  galaxy ID 5519,  allows us to present the hypothesis that at least some of such rings could be remnants of relic magnetic wormholes. 
It turns out that magnetic wormholes may directly collect ring clumps of baryons in a process which is not based on the development of perturbations in dark matter component. This means that the dark matter halo in such structures may have peculiar features, different from those observed in galaxies, e.g. see  \cite{PSS,Sal16} and references therein. In particular, in the case of wormholes the dark matter halo can be almost absent. This  argument however does not work, if dark matter phenomenon has a pure topological origin \cite{KT},  or it appears due to some extension of general relativity  \cite{K06}.  The only stringent feature of such objects is the presence of a large - scale toroidal magnetic field.
In the rest, the final form of the ring galaxy formed by a wormhole does not differ from that predicted by standard mechanisms. Indeed, upon the recombination epoch in the case when the wormhole collapses the ring may contain a more dense and old collection of stars (a small galaxy) in the center of the ring (as in the case of the Hoag's object), or, when the wormhole expands further, the center part of the ring may remain empty, e.g., see examples  in  \cite{R3,ER}.

Wormholes are exotic objects whose existence is predicted by general relativity (GR), e.g. see the history in Refs. 
  \cite{Vis}. Enormous efforts have been made to construct exact solutions of the Einstein equations which correspond to a stable wormhole, e.g., see some examples in Refs \cite{wh}. The dry rest is that in the case of  spherically symmetric configurations  wormholes are highly unstable and collapse very rapidly. To be stable they require the presence of exotic (violating the weak energy condition) matter. 
This means that all primordial spherical wormholes have collapsed long ago and they cannot be distinguished from black holes\footnote{The only difference is that remnants of wormholes may contain magnetic poles.}. 

Over the last decade there is an essential increase of interest in different modifications of GR in which the violating energy condition matter is replaced by an appropriate modification e.g., see Refs \cite{Sushk,Sushk13,Vagnozzi}. We leave aside such a 
possibility and explore less symmetric configurations. 
It turns out that less symmetric configurations can be made stable without exotic matter or any modification of GR.  First rigorous example was presented in Ref. \cite{KS16}. It was demonstrated that in the open Friedman model a stable wormhole can be obtained simply by the factorization of space over a discrete subgroup of the group of motions of space\footnote{To illustrate such a factorization one may consider flat space in which one of coordinates, say $x$, becomes periodic $x=R\varphi /(2\pi )$ with the angle $0<\varphi <2\pi $. Then the space becomes a cylinder with the radius $R$. In the open model (on space of a constant negative curvature) any parallel geodesics ($\varphi =0$ and $\varphi =2\pi $) diverge and the distance between them changes $R(\ell)$, where $\ell$ is a parameter along the geodesic line. Therefore, if we move along the geodesic line from the point where the distance is the shortest $R_{min}$, the space opens out $R\rightarrow \infty$ and becomes unrestricted. This is illustrated on Fig. 1. in Ref. \cite{KS20}. In other words, the simplest stable wormhole corresponds simply to a torus on the Lobachevsky space. }. We point out that such a procedure does not allow the spherical configurations at all. 
The simplest wormhole obtained by the factorization has the throat in the form of a torus, i.e., it has the shape of a doughnut  \cite{KS20}.  We stress that such  wormholes are not static but expand in agreement with the expansion of the Universe. They are frozen into space and, therefore, are static in the co-moving coordinates.  It was also demonstrated that the factorization allows to get an arbitrary number of such wormholes in space. 

In flat space the torus - like wormholes become dynamical objects and evolve \cite{KS16}. Whether they are static, expand, or collapse, depends on surrounding matter and peculiar motions of throats. The shape of throats of such wormholes resembles a doughnut and is characterized by two radii $R_w$ and $r_w$. In the limit $R_w\gg r_w$ it can be approximately described by the cylindrical (axial) 
configuration. It turns out that static and stationary cylindrical wormhole solutions do exist and it was found that asymptotically flat wormhole configurations do not require exotic matter violating the weak energy condition \cite{Bron6,Bron8}. Such encouraging results show that such objects, as doughnut - shaped wormholes, have all chances to be observed in astrophysical systems. We stress that the complete description of the evolution of such wormholes in flat space represents a rather complex problem which still awaits for the rigorous investigation.

The possibility to  directly observe  wormholes attracts the more increasing attention, e.g., see Refs. \cite{Ghersi:2019trn,lens1,lens2,lens41}. At first glance the most promising are collective effects  produced by a distribution of wormholes in space. However, our previous  investigation have shown that observational effects of a distribution of wormholes are very well hidden under analogous effects produced by ordinary matter e.g., see Refs. \cite{KS18,KS19}. The only exclusion may be the noise (stochastic background) produced by the scattering of emitted by binaries gravitational waves  on wormholes \cite{KSL20}.  

In general 
a single wormhole produces much less noticeable effects (lensing,  cosmic ray scattering, etc.).
However,  
wormholes may possess non-trivial magnetic fields as vacuum solutions. In this case possible imprints of wormholes in the present picture of the Universe may be rather considerable.
In particular, when such a magnetic wormhole gets close to a galaxy, it starts to work as an accelerator of charged particles \cite{KS20} which  is capable of explaining the origin of high-energy cosmic-ray particles \cite{rays}. 
The idea  that  the observed cosmic rays require magnetic fields for their creation
 was first suggested by Fermi \cite{F49}.
In voids a relic magnetic wormhole works simply as a generator of synchrotron radiation and can be detected via the magnetic field \cite{MF1,MF2}. The primordial magnetic fields \cite{MF3} in turn may form small-scale non-linear clumps of baryonic matter \cite{BCF,BCF2} and as it was recently shown  \cite{MFG3}  they allow to solve the existing  tension between the Hubble constant value measured by 
 Planck  $H_0=67.36\pm 0.54 km \ s^{-1} Mpc^{-1}$ \cite{Planck18} and measured by the Supernovae $H_0 = 74.03 \pm 1.42 km s^{-1} Mpc^{-1}$ 
 \cite{SH0}.
In other words, magnetic wormholes should leave a clear imprint on the sky. It turns out that relic magnetic wormholes may play also the key role in formation of ring type baryonic structures, analogous to the ring galaxies.

\section{Magnetic wormholes as baryon traps}
Consider a single wormhole whose throat has the shape  of a torus (the genus -$1$ wormhole by the classification suggested in Ref. \cite{KS20}). In the presence of such a wormhole Maxwell's equations possess two additional classes of non-trivial  vacuum solutions. 
Indeed, according to the Stocks theorem
the system of vacuum Maxwell equations implies $\oint \mathbf{Bdl}=0$ for any loop  which can be pulled to a point (where $\mathbf{B}$ is the magnetic field).
In the case of a non-trivial topology of space\footnote{The simplest scheme to get a general  genus - n wormhole can be described as follows. We take a couple of equal spheres with $n$ handles in space, remove the internal regions (insides of the spheres), and glue along their surfaces (the so-called Hegor diagrams). If we take two simple spheres without handles the resulting space corresponds to a spherical wormhole (throat is the sphere). The sphere with a handle is the torus. A couple of toruses corresponds to the doughnut shaped wormhole, etc.. } there appear new classes of loops $\Gamma_a$ which cannot be contracted to a point and, therefore, to fix the unique solution we have to fix additional boundary data $\oint _{\Gamma_a}\mathbf{Bdl}=\frac{4\pi }{c}I_a$ and in general $I_a\neq 0$. The constants $I_a$ depend only on time and they can be viewed as fictitious currents\footnote{We point out  that the currents are fictitious, for from the point of view of the Hegor diagrams they take place in inner regions of  spheres which are removed, i.e., in fictitious regions.} 
 which intersect the loops $\Gamma_a$. 
In the case of genus $n=0$ (spherical) wormhole there is only one such a non-trivial loop which goes through the wormhole throat. In the case of genus $n=1$ wormhole (doughnut - shaped throat) we have already two such loops, one goes through the throat  and one additional goes through the hole in the center of the doughnut and surrounds the throat.

The first class produces the magnetic field of a wormhole which can be described by  magnetic poles placed in  two different entrances into the throat.  The two entrances have opposite magnetic poles. If the distance between the entrances is big enough the resulting field is very weak and when crossing such a field high-energy charged particles only slightly change the direction of propagation. The field can be strong only very close to the throat entrances. However, since charged particles can freely  propagate along the field lines
(which are roughly orthogonal to entrances), the particles captured by the field are distributed in the whole region between the entrances. Such fields have the long-range character and can be used to explain the origin of long-correlated magnetic fields in voids \cite{MF1,MF2} and, more generally, of primordial magnetic fields \cite{MF3}.

The situation changes when the wormhole possesses also the field of the second class.
The second class corresponds to the field produced by a single loop of a current (the loop of the corresponding fictitious current goes inside of the surface of the doughnut). In this case the field lines of force repeat the shape of the entrance (the shape of a doughnut) which roughly corresponds to the fields observed in spiral galaxies \cite{MFGal,MFG2,MFG3}. 
It is necessary to point out that the magnetic fields in the galaxies cannot exactly repeat the shape of a doughnut. There are a number of well-known processes in galaxies which naturally lead to the generation of galactic magnetic fields \cite{MFG2} and the actual field does not reduce to such a simple model. It is known that the pitch angle (between the toroidal field and radial one) is usually between 4 and 17 degrees.

The most strong field is close to the entrance and particles captured by the field remain always near the entrance.
In the primordial plasma before the recombination such a wormhole traps all baryons\footnote{We present here estimates for baryons only, since leptons are much lighter than baryons and in the primordial plasma leptons simply follow baryons (bounded by the Coulomb potential).} propagating near it and thus forms a primeval structure of a galaxy (a protogalaxy).
Such a scheme works only for wormholes which possess sufficiently strong magnetic fields, since high-energy  baryons cannot be captured by the wormhole. The mean energy of baryons is determined by the temperature which depends on the redshift. The intensity of the magnetic field also depends on the redshift. While baryons are relativistic particles 
the threshold  value of the fictitious (or equivalent) current does not depend on time.

Indeed, only particles below the threshold energy are captured by the wormhole magnetic field   \cite{KS20}   which is given by
\begin{equation*}
E=3kT<E_{th}=eBR_w,
\end{equation*}
where $e$ is the electron charge and $R_w$ is the biggest radius of the doughnut - shaped throat of the wormhole. The energy of relativistic baryons behaves with the redshift as $T=T_{\gamma }(1+z)$, where $T_{\gamma }$ is the present day temperature of CMB radiation.
The intensity of the magnetic field can be estimated by the value in the hole of the doughnut. We take it as
\begin{equation*}
B= \frac{\kappa  I}{cR_w},
\end{equation*}
where $\kappa =2\pi $ in the center of the doughnut hole, while 
 close to the throat (surface of the doughnut),  where the field reaches the maximum value, in the approximation $ r_w/R_w\ll 1$ we get the estimate $\kappa \simeq 2R_w/r_w$.
The field depends on the  parameter $I$ (which is the fictitious or equivalent current) and the big  radius  $R_w$ of the doughnut which also depends on the redshift as $R_w=R_0/(1+z)$. 

The constant $I$ behaves with the redshift as $I=I_0(1+z)$. Indeed the invariant characteristics is the number of magnetic lines captured by the throat (which go through the internal hole of the doughnut). This gives $\int _S\mathbf{Bds} = const=\Phi $, where $S$ an arbitrary surface whose  boundary contour $\gamma$ (dual to $\Gamma $) lays on the surface of the throat and cannot be contracted to a point, so that all magnetic lines intersect $S$ only once. Taking the minimal surface $S$ we find $\Phi \sim  \frac{2\pi ^2 }{c} R_wI $ which gives the behavior $I\sim 1/R_w\sim 1/a$, where $a=a_0/(1+z)$ is the scale factor of the Universe. We point out that the same dependence on the redshift $z$  follows from the fact that the energy density of the magnetic field $\rho _B \sim B^2/4\pi $  decreases with the scale factor as $\rho _B\sim 1/a^4$.
This gives the threshold value for the equivalent current which defines the intensity of the magnetic field as 
\begin{equation}
I_0> I_{th}  = \frac{ce}{\kappa r_{\gamma }}  \sim  \frac{3.2}{\kappa }\times 10^{-5} A ,
\end{equation}
where 
$r_{\gamma}=\frac{e^2}{3kT_{\gamma }}
$. It is convenient to express $r_{\gamma}$ as follows $r_{\gamma}=r_p(1+z_r)$, where 
$r_p=\frac{e^2}{m_p c^2}
$ is the classical radius of the  proton and 
 $
1+z_r =\frac{m_p c^2}{3kT_{\gamma}} \sim 10^{12}
$ is the redshift at which baryons become relativistic particles.
It is curious that the threshold value $I_{th}$ is extremely small. It does not depend on the absolute size of the wormhole (which is given by the big radius $R_w$) but only on the ratio of the wormhole radii $\kappa =2R_w/r_w$. 
All wormholes with the present day values  $I_0>I_{th}$ strongly interact with baryons. We may say that they are frozen into baryons and, therefore, peculiar motions of baryons repeat peculiar motions of such wormholes. Wormholes with smaller magnetic fields $I_0<I_{th}$ slightly interact with baryons and can be considered as free objects. They may participate in independent from baryons motions. As we shall see on the early stage of the evolution of the Universe, before the recombination, they cannot capture baryons and therefore do not form an enhancement in the baryon density. On latter stages upon reheating they may capture charged particles and form cosmic rays and compact sources of synchrotron radiation. The relativistic stage for baryons  finishes at the redshift 
$ z_r$ upon which (e.g., for $z<z_r$) baryons become non-relativistic and the above consideration brakes. 

At redshifts  $z_r>z>z_{rec}$ baryons are non-relativistic, while upon the recombination $z_{rec}$ baryons form neutral Hydrogen atoms and do not interact with magnetic fields of all wormholes. They again start to interact with wormholes only at the epoch of re-ionization when first stars have fired. 

On the stage $z_r>z>z_{rec}$  for every particular wormhole it is possible to define the critical redshift $z_0$ when it captures baryons (for $z>z_0$). The critical value $z_0$ can be estimated as follows.
The mean energy of baryons is $m_pV^2/2=3kT/2$ with $T=T_{\gamma}(1+z)$. 
Then the critical redshift can be found from the inequality $r_B=\frac{V}{\omega _B}=\frac{Vm_pc}{eB}< r_d$, where $r_B$ and $\omega _B$ are Larmor radius and frequency respectively for protons, while $r_d$ denotes the width of the baryon cloud around the wormhole throat. The width $r_d$ should be smaller, than the proton diffusion length and it defines the critical value of the magnetic field. 
This gives 
\begin{equation*}
B=\frac{\kappa  I}{cR_w} > \frac{m_pc}{er_d} \sqrt{\frac{3kT}{m_p}},
\end{equation*}
or equivalently 
\begin{equation*}
  I_0(1+z)> \frac{\alpha e}{ \kappa r_p} \sqrt{\frac{3kT}{m_p}}=  \alpha I_{th}(1+z_r)\sqrt{\frac{(1+z)}{(1+z_r)}},
\end{equation*}
where the ratio $\alpha =\frac{R_w}{r_d}\gg 1$.
This defines the critical redshift $z_0$ at which the wormhole starts to trap baryons as
\begin{equation*}
(1+z)> (1+z_0)=\alpha ^2\frac{ I^2_{th}}{  I^2_0}(1+z_r).
\end{equation*}
The relation between $z_0$ and $I_0$ can be rewritten as 
\begin{equation}\label{z_0}
 I^2_0= \frac{ (1+z_r)}{ (1+z_0)}\alpha ^2I^2_{th}.
\end{equation}
When $z_0=z_r$ we get $I_0=\alpha I_{th}$ and if we take $\alpha =1$, this will give the absolute threshold of the field. We point out that 
the value $\alpha =1$ may have sense only for sufficiently small wormholes with the radius $R_w\lesssim10^{-5}R_{gal}$, e.g., see estimates in the next section.
At redshifts $z>z_r$ we get into the epoch where baryons are relativistic particles and wormholes with $ I_0<I_{th}$ do not bound baryons at all.
There is one more critical value $I_{rec}$ which corresponds to  $z_0=z_{rec}$.
\begin{equation}
 I^2_{rec} =\alpha ^2I^2_{th}  \frac{(1+z_r)}{(1+z_{rec})} \gg  I^2_{th}.
\end{equation}
 All fields with $I_0>I_{rec}$ are strong enough to capture baryons during the whole evolution  $z>z_{rec}$. 
 
 \section{The number of baryons in traps}
 The efficiency of wormhole traps can be described by the number of baryons collected.
The number of baryons collected around magnetic wormholes depends on the proton diffusion length $\ell (z)$,  the big radius  a wormhole throat $R_w(z)$, and the radius of the baryon cloud $r_{cl}(z)\ll R_w$.  For estimates one may use $r_{cl}\sim r_w$. This number can be estimated as the increase of the effective volume of the torus-shaped throat  
$$
\Delta N=<n_b> \left(V(R_w+\ell ,r_{cl}+\ell)-V(R_w,r_{cl})\right) ,
$$
where  $<n_b>$ is the mean density of baryons and $V=2\pi^2R_wr_{cl}^2$ is the throat volume. We define the parameter $\delta _b =\Delta N/(V<n_b>) $ 
which depends on the position in space. Close to wormhole throats $\delta _b>0$, while sufficiently far from the wormhole $\delta _b<0$ since baryons from those regions have captured by the wormhole. The value $<\delta _b^2>=b$, where brackets define the averaging  over the space,  relates to the baryon clumping factor   $b=(<n_b^2>-<n_b>^2)/<n_b>^2$, e.g., \cite{BCF,BCF2}.

In the case $\ell (z)\ll r_{cl}(z)$ we get 
 \begin{equation}\label{y1}
 \delta _b(z) \sim  \frac{\ell (z)}{R_w(z)} +\frac{2\ell (z)}{r_{cl}(z)}\ll 1.
\end{equation}
In the intermediate case  $ r_{cl}(z)\ll \ell (z)\ll R_w(z)$ we find the estimate 
\begin{equation}\label{y02}
 \delta _b(z) \sim   \frac{\ell ^2(z)}{r^2_{cl}(z)}(1+\frac{\ell (z)}{R_w(z)})\gg 1.
\end{equation}
And in the case $  \ell (z)\gg R_w(z)$ the estimate has the order 
\begin{equation}\label{y03}
 \delta _b(z) \sim   \frac{\ell ^3(z)}{R_w(z)r^2_{cl}(z)}\gg 1.
\end{equation}

On the  stage $z>z_r$ protons are relativistic particles, plasma is degenerate, and the length of the proton propagation can be estimated by the value of the horizon size 
 $\ell (z)\sim l_h=c/H(z)$. At the redshift $z=z_r$ it is extremely small and has the order $\ell (z_r)\sim (7 \div 8)\times 10^{-15} pc$. Consider a wormhole throat with the big radius $R_w(0)\sim 15kpc$ which corresponds to a galaxy size \cite{R3}, while the radius of the baryon cloud has the order $r_{cl}\sim 0.2R_w$. Then we find $R_w(z_r)\sim 15\times 10^{-9} pc\gg \ell(z_r)$ and from  (\ref{y1}) we get 
 $ \delta _b(z_r) \sim (0.5 \div 0.6)\times 10^{-5} $. Such a value is too small to form a galaxy without additional means (e.g., a dark matter clump).

 At the recombination $z_{rec}=1100$ the proton diffusion length  has the co-moving value of the order $\ell (z_{rec})\sim 0.4\div 1pc$, while $R_w(z_{rec})\sim 13.6pc$ and we still may consider $\ell (z_{rec})\ll R_w$ and $\ell \lesssim r_{cl}$. Therefore we again may use (\ref{y1}) and find 
 $  \delta _b(z) \sim  0.32\div 0.8 $. Such a big value shows that the respective protogalaxy forms immediately after the recombination and it is already in the non-linear regime.
We should expect that  wormholes with  such strong clumps of baryons depart the Hubble expansion very soon and form rather small objects.
 Smaller wormholes form too strong inhomogeneities before recombination and probably collapse to blackholes. We point out that this may give a new mechanism of blackhole formation with huge masses. 
 
 Galaxies  are observed up to $z\simeq 11$ (e.g., the most distant GN-z11 is observed at $z=11.09$ \cite{GN-z11}). Therefore, to be consistent with the present day size of a typical ring galaxy the wormhole radius should be at least two orders bigger $R_w(0)\lesssim 1Mpc$, which gives already $R_w(z_{rec})\lesssim 900 pc$ and $  \delta _b(z_{rec}) \sim  4,8\times 10^{-3}$. Such a clump departs the Hubble expansion already at $z\sim 100$ and gives a ring of the order $R\sim 10 Kpc$.


 \section{Conclusions}
 In conclusion we point out two important facts. First one is that the wormholes whose size $R(z_{rec})$ exceeds the value $\ell (z_{rec})$ more than on the factor $10^3$  do not form a sufficient enhancement in the baryon number density and, therefore, cannot form ring galaxies. 
 They however may form ring - type structures in the future, e.g. see the recently reported findings of unexpected class of astronomical ring-type objects in  \cite{ORCs}.

 The second fact is that upon the recombination $z<z_{rec}$ the doughnut - shaped wormholes do not interact with baryons and evolve. Therefore, they may leave the ring clump formed. They either expand or collapse forming a magnetized blackhole in the middle. 
 If the wormhole collapses, it should also draw some portion of the baryon clump and form a bulge in the center of the ring. In this case it may form the ideal symmetric structure similar to the Hoag's object.  Additional rotational perturbations of clumps may however lead to irregular structures.
  If the wormhole expands further, the center part of the ring remains to be empty.  We may expect that in such a case
 the mean value of the magnetic field within the ring retains.  We recall that different active processes in galaxies generate magnetic fields which in general have a turbulent character. Therefore, by the measuring the mean flow of the magnetic field in a ring galaxy one may at least estimate the present day value of the big radius $ R=\frac{\kappa  I}{cB}$  of such a wormhole and estimate its present day position. 
This may be used in the direct search  for wormhole traces in the Universe. 

At first glance the most direct indication on the possible role of relic magnetic wormholes in the formation of some ring galaxies should be the discrepancy between the observed amount of dark matter in such galaxies and the predictions of the standard theory  \cite{Hierarchy1,Hierarchy2,PSS}. However there exist some extensions of general relativity which are capable of reproducing dark matter effects in galaxies without dark matter particles, e.g., see \cite{KT,K06} and references therein.
Therefore, we think that the only rigorous indication on the presence of relic magnetic wormholes should be large-scale toroidal magnetic fields.

\section{Acknowlegment}
We acknowledge valuable comments and the advice of referees which helped us to essentially improve the presentation of this work. 


\end{document}